\documentclass[prd,a4paper,aps,twocolumn,showpacs,
               nofootinbib,nobibnotes,superscriptaddress]
              {revtex4}
\usepackage{epsfig}
\usepackage{amssymb}

\newcommand{\Ga}{\Gamma}

\newcommand{\la}{\lambda} 
\newcommand{\Om}{\Omega}

\newcommand{\be}{\begin{equation}} 
\newcommand{\ee}{\end{equation}} 
 
\newcommand{\lsim}{\lesssim} 
\newcommand{\bea}{\begin{eqnarray}} 
\newcommand{\eea}{\end{eqnarray}} 
\newcommand{\bean}{\begin{eqnarray*}} 
\newcommand{\eean}{\end{eqnarray*}}

\newcommand{\mr}{\mathrm }

\newcommand{\Gauss}{{\mbox{Gauss}}}

\begin{document} 

\title{Limits on stochastic magnetic fields: A defense of our paper
  \protect\cite{ours}}  

\author{Chiara Caprini} 
\email{chiara.caprini@physics.unige.ch} 
\affiliation{D\'epartement de Physique Th\'eorique, Universit\'e de 
  Gen\`eve, 24 quai Ernest Ansermet, CH--1211 Gen\`eve 4, Switzerland} 
 
\author{Ruth Durrer} 
\email{ruth.durrer@physics.unige.ch} 
\affiliation{D\'epartement de Physique Th\'eorique, Universit\'e de 
  Gen\`eve, 24 quai Ernest Ansermet, CH--1211 Gen\`eve 4, Switzerland} 

\date{\today} 
 
\begin{abstract} 
 In their recent paper ``Faraday rotation of the cosmic microwave background
polarization by a stochastic magnetic field'', Kosowsky \emph{et
  al}.~\cite{Koso} have commented about our
 paper~\cite{ours}, in which we derived very strong limits on the amplitude of
 a primordial magnetic field 
 from gravitational wave production. They argue that our limits are erroneous. 
In this short comment we defend our result.

\end{abstract}

\maketitle

In Ref.~\cite{ours} we have shown that, if a magnetic field is present on
super-horizon scales in the early universe, its power 
is very efficiently converted into gravitational waves during
its evolution from super- to sub-horizon scales. We used this fact to
derive stringent limits on the amplitude of a magnetic field created before
the nucleosynthesis epoch. 

In their recent paper~\cite{Koso}, Kosowsky \emph{et al}. state that
our limits are not valid. In the discussion section, 
they argue ``... the expansion rate of the
universe is the same whether energy density is converted from magnetic
fields into gravity waves or not, since the energy density of both
scale the same way with the expansion of the universe. So the actual
constraint is on the total radiation energy density in the magnetic
field, which is constrained to be about 1\% of the total energy density
in the usual manner... The corresponding limit on the total comoving
mean magnetic field strength is around $10^{-8} \Gauss$, not the
$10^{-27}\Gauss$ claimed in~\cite{ours}.''

We now explain why this conclusion is wrong. We employ the same
notation convention as~\cite{ours}, and we
always consider the comoving amplitude of the magnetic field. 
For magnetic
fields with spectral index $n>-3$, the magnetic field energy at wave 
number $k$ is given by 
 \be
{d\Om_B(k)\over d\log(k)} = {B_\la^2\over 8\pi\rho_c}{(k\la)^{n+3}\over
2^{(n+3)/2}\Ga({n+3\over 2})} \label{OmBk}~,
\ee
where $B_\la$ is the magnetic field amplitude at some fixed reference
scale $\la$.
This energy spectrum is always blue, and therefore dominated by its
value at the upper cutoff $k_c$. This cutoff scale is time
dependent, $k_c(\eta)$. We
set the magnetic field to zero on scales which are already sub-horizon
at the time $\eta_*$ of formation of the magnetic field, 
since we cannot be sure that its spectrum is a power law on
these very small scales. Therefore, the
upper cutoff at the time of formation
of the magnetic field is given by $k_*=\eta_*^{-1}$ (where $\eta$ denotes
conformal time). This assumption is a conservative one for the derivation of
our result. 
At later times, the magnetic field is damped
on scales smaller than a time dependent damping scale, which gives us the
cutoff $k_D(\eta)$~\cite{Subra,Jeda}.
We therefore obtain the cutoff function 
\be k_c(\eta)   =\min\left(    k_*,k_D(\eta)\right)~.
\ee
Of course at formation $k_* =1/\eta_* \ll k_D(\eta_*)$, while
at later time $ k_D(\eta)$ is decreasing, and eventually becomes smaller
that $k_*$. The magnetic field energy density at a given 
time $\eta$ is therefore given by
\bea
\Om_B(\eta)= \Om_B(k_c(\eta)) &=& 
\int_0^{k_c(\eta)}{dk\over k}{d\Om_B(k)\over d\log(k)}
\nonumber\\
  &=& {B_\la^2\over 8\pi\rho_c}
{(k_c(\eta)\la)^{n+3}\over 2^{(n+5)/2}\Ga({n+5\over 2})}~.
\label{OmB}
\eea
In our paper~\cite{ours}, we have shown that at the time a given scale
crosses the horizon, and for the maximally allowed magnetic fields
($\Om_B \sim \Om_\mr{rad}$, see Eqs.~(24) and (26) in
Ref.~\cite{ours}), a considerable fraction\footnote{For some values of
  the spectral index we obtain more energy in gravity waves than 
  in the magnetic field. This comes from the fact that
  we linearize the problem and therefore do not take into account
  backreaction. We expect the correct fraction of the energy in
  gravity waves to lie between 50\% and 100\% of the magnetic field
  energy density.} of the magnetic field energy density is 
converted into gravitational wave energy density:
$$ \left. \frac{d\Om_{G}(k,\eta)}{d\log(k)}  \right|_{k=1/\eta} \sim  
    ~~~ \frac{d\Om_{B}(k)}{d\log(k)} ~.$$
As time goes on, the magnetic field is damped on sub-horizon scales
$k> k_D(\eta)\gg 1/\eta$, while the gravitational waves are not damped,
  since after formation they no
longer interact with the matter and the radiation in the universe. This is
the main point which Kosowsky \emph{et al}. have missed. 
The magnetic field density parameter at nucleosynthesis is given by
\be
\Om_B(\eta_\mathrm{nuc}) = {B_\la^2\over 8\pi\rho_c}
{(k_D(\eta_\mathrm{nuc})\la)^{n+3}\over 2^{(n+5)/2}\Ga({n+5\over 2})}~,
\label{OmBnuc}
\ee
where we have integrated up to the cutoff at nucleosynthesis, $k_D(\eta_{\rm
  nuc})$; while the gravitational wave density parameter is
\be
\Om_G \simeq \Om_B(\eta_*) = {B_\la^2\over 8\pi\rho_c}
{(k_*\la)^{n+3}\over 2^{(n+5)/2}\Ga({n+5\over 2})}~,
\label{Omg}
\ee
where we integrate up to the cutoff corresponding to the time of creation of
the magnetic field: gravitational wave production for a magnetic mode $k$ 
takes place before horizon
crossing, before the magnetic field is damped by interaction with the cosmic
plasma. A considerable part of the magnetic energy is therefore
``stored'' in gravitational waves. 

From (\ref{OmBnuc}) and (\ref{Omg}), one can see that the ratio
between the two energy densities is $\Om_B(\eta_\mathrm{nuc})/\Om_G
\simeq 
(k_D(\eta_\mathrm{nuc})/k_*)^{n+3}$. Only for $n\simeq -3$ or $\eta_*
\simeq \eta_\mr{nuc}$ this factor is of order unity; in this case, one can
apply the nucleosynthesis bound indifferently to $\Om_G$ or $\Om_B$, and one
gets the same constraint. In most cases instead, 
the ratio is rather huge. Let us
consider the example of magnetic field generation at the electroweak
phase transition. In~\cite{ours} we calculate $k_D(\eta_\mr{nuc}) \simeq
6\times 10^{-7}$sec$^{-1} \sim 60$~pc$^{-1}$, and $\eta_\mr{ew} \simeq 4\times
10^4$~sec. Taking into account that electroweak magnetic field
generation is causal (not inflationary), and therefore n=2
(see~\cite{causal}), we obtain
\be
\left. \frac{\Om_B(\eta_\mathrm{nuc})}{\Om_G}\right|_{\eta_*=\eta_{\rm ew}} 
\simeq (k_D(\eta_{\rm nuc})\eta_{\rm ew})^5
\simeq 8\times 10^{-9}~, 
\ee
and by no means one! If the magnetic field is generated during
inflation, one is no longer forced to have $n=2$, but can have
arbitrary values of $n>-3$. If we take $n \simeq 0$,
for an inflation scale of $10^{15}$~GeV, we have $k_* =1/\eta_\mr{inf} \simeq
10^{13}/\eta_\mr{ew}$, and we obtain 
\be
\left. \frac{\Om_B(\eta_\mathrm{nuc})}{\Om_G}\right|_{\eta_*=\eta_{\rm ew}} 
\simeq (k_D(\eta_{\rm nuc})\eta_{\rm inf})^3
\simeq 10^{-43} ~~ !!
\ee
We can conclude that, applying the nucleosynthesis bound on $\Om_G$, we find
much stronger constraints on $B_\la$, due to the fact that a seizable
fraction of the magnetic field energy is converted into gravitational
waves before the damping process.  

Apart from not taking into account this damping, there is a second point that
has been missed in Kosowsky \emph{et al}. 
We specifically pronounce a limit for the amplitude of the stochastic 
magnetic field smoothed over a scale $\lambda
\sim 0.1$~Mpc. On the contrary, they talk about `the total comoving mean magnetic
field', which is largely dominated by its value on small scales, hence
$B(k_D(\eta_\mr{nuc}))$. The value of the field at this scale is limited
to a few $10^{-8}\Gauss$ by the constraint $\Om_B <
0.1\,\Om_\mr{rad}$ at nucleosynthesis. But this field value has
no relevance, for two reasons. First of all, because the damping scale will
grow, which means that $B(k_D(\eta_\mr{nuc}))$ will be damped away
before it can ever give rise to magnetic fields in galaxies. Typically, the
highest mode which survives damping is $k_D(\eta_{\rm rec})\simeq 10 \,{\rm
  Mpc}^{-1}$, much smaller than $k_D(\eta_{\rm nuc})$~\cite{Subra,Jeda}.  
Secondly, the scale relevant for magnetic fields in galaxies and clusters is
$\la\sim 0.1$---$1$~Mpc, and we have thus formulated limits for this scale. 
 If $B(k_D(\eta_\mr{nuc}))k_D (\eta_\mr{nuc})^{3/2} \equiv B_{k_D}
 \lsim 10^{-8}\Gauss$,   
the limit on the scale $\la\gg 1/k_D(\eta_\mr{nuc})$ is much smaller,
namely $B_\la = B(k=1/\la)\la^{-3/2} =B_{k_D}(k_D
  (\eta_\mr{nuc})\la)^{-(n+3)/2} \lsim 10^{-8}\Gauss\times 10^{-6(n+3)/2}$. 
For a spectral index $n=2$, for example, $B_\la$ is
smaller than the maximal field  by a factor of $\sim 10^{15}$,
namely $B_\la \lsim 10^{-23}\,\Gauss$.

In conclusion, even if the magnetic field and gravitational wave energy
densities scale in the same way with the expansion if the universe, applying
the nucleosynthesis bound on the induced gravitational wave energy density
gives a much stronger constraint on the amplitude of the magnetic field. Both
the magnetic field and gravitational wave energy spectra are blue, and
therefore dominated by their value at the upper cutoff. However, the upper
cutoff for the gravitational wave spectrum is much higher than the one for
the magnetic field spectrum at the epoch of nucleosynthesis: $\eta_*^{-1}\gg
k_D(\eta_{\rm nuc})$. The reason for this being, that the conversion of
magnetic field energy into gravitational wave energy takes place when a given
mode enters the horizon, before the magnetic field is dissipated by
interaction with the cosmic fluid. 

Furthermore, the interesting limit
is not the one on the `mean magnetic field' which is dominated by the
value at the smallest scale, but the limit on the field
amplitude at some scale $\la$ which is relevant for galactic magnetic
fields, and certainly has to be larger than the damping scale at the
redshift of galaxy formation.

Of course, these limits apply for magnetic
fields generated before the epoch of nucleosynthesis.  

\acknowledgments
We acknowledge intensive and  stimulating discussions with Tina
Kahniashvili and Arthur Kosowsky.

\end{document}